\documentclass[11pt, letterpaper]{article}

\usepackage[utf8]{inputenc}
\usepackage[T1]{fontenc}
\usepackage{amsmath, amssymb}
\usepackage{graphicx}
\usepackage{geometry}
\usepackage{authblk}
\usepackage{booktabs}
\usepackage{rotating}
\usepackage{tikz}
\usepackage[numbers, sort&compress]{natbib}
\usepackage{xcolor}
\usepackage[colorlinks=true, linkcolor=blue, citecolor=blue, urlcolor=blue]{hyperref}

\title{RevealLO: Global-Scale FWI Including Long-Period Major-Arc Surface Wave and Multi-Orbit Body Wave Data}

\author[1]{Carl Josef Schiller\thanks{Corresponding author: \href{mailto:carl.schiller@eaps.ethz.ch}{carl.schiller@eaps.ethz.ch}}}
\author[1]{Scott Keating}
\author[2]{Michael Afanasiev}
\author[2]{Patrick Marty}
\author[1]{Andreas Fichtner}

\affil[1]{Department of Earth and Planetary Sciences, ETH Zurich, Zurich, Switzerland}
\affil[2]{Mondaic Ltd., Zurich, Switzerland}

\date{\today}

\begin{document}

\maketitle


\begin{abstract}
Global full-waveform inversion (FWI) has matured significantly in recent years, but resolution in the southern hemisphere often remains limited due to sparse receiver coverage. To address this problem, we constructed RevealLO, a global-scale FWI model that updates its predecessor REVEAL by expanding the inversion time window from 1 h to 3.5 h. The extended time window incorporates both minor-arc ($R_1, G_1$) and major-arc ($R_2, G_2$) surface wave arrivals, as well as late-arriving multi-orbit body waves, thereby increasing azimuthal coverage beneath ocean basins and the southern hemisphere. We simulate 3-D viscoelastic wavefields using the spectral-element solver Salvus on global fourth-order cubed-sphere meshes, and update velocity structure by successively broadening the period band from 100 - 200 s to 60 - 200 s. Measurements and adjoint sources are based on time-frequency phase misfits, which are suitable for both large-amplitude surface and low-amplitude multiply-reflected body waves. For iterative optimization, we employ a mini-batch stochastic trust-region L-BFGS algorithm, followed by mesh reprojection onto the 33~s REVEAL baseline mesh and lowermost-mantle regularization. RevealLO preserves short-period whole-Earth body wave fits while improving azimuthal sampling and mitigating trade-offs between unmodeled azimuthal anisotropy and isotropic structure in the upper and mid mantle.
\end{abstract}


\section{Introduction}
\label{sec:intro}

Over the past decade, full-waveform inversion (FWI) has developed into a high-resolution method for imaging the Earth's interior. By exploiting complete waveform segments rather than isolated travel-time picks (\cite{Tarantola1984}; \cite{Pratt1999}), FWI accounts for complex wavefield phenomena such as multipathing, wavefront healing (\cite{Wielandt1987}; \cite{Nolet2000}), and finite-frequency sensitivity (\cite{Marquering1999}, \cite{Dahlen2000}, \cite{Tromp2005}). Although current global tomographic models exhibit agreement at long wavelengths, structural discrepancies persist at shorter scales \cite{Fichtner_2025b}. These differences are primarily driven by data selection choices, phase windowing, parameterization, and inversion strategies employed by different research groups.\\[5pt]
REVEAL was recently developed as a global-scale FWI model utilizing a multi-scale FWI strategy \cite{thrastarson2024}. To mitigate non-linearity, improve convexivity, and avoid cycle-skipping, the iterative construction of REVEAL progressed from long to short periods, ending at a minimum period of 33~s. The maximum length of the seismograms was set to 1~hour, mostly due to cost constraints. While this approach includes direct body waves and minor-arc surface waves, it omits potentially valuable information contained in later arrivals.\\[5pt]
Specifically, truncating seismograms at 1~hour excludes surface waves traveling along major-arc paths, thus limiting upper-mantle sensitivity in the southern hemisphere and beneath ocean basins, where seismic station density is lower than in the northern hemisphere. To illuminate these regions, it is necessary to take advantage surface and body waves that sample the Earth at epicentral distances that exceed $180^\circ$.\\[5pt]
To close this gap, we performed a long-period/long-duration update of REVEAL, leading to \mbox{RevealLO}. The primary modification in the construction of RevealLO is the incorporation of 3.5~hour seismograms, which include major-arc ($R_2, G_2$) surface waves, multi-orbit body waves, and few multi-orbit surface waves (e.g., $G_3$). Further multi-orbit arrivals are drowned in the noise of the considered period bandpass, thus we truncate to 3.5~hours for one full orbit of the wavefield. The main effect is to increase the coverage of the southern hemisphere upper mantle and transition zone.\\[5pt]
Starting from a regularized version of REVEAL model, we revise the long-period structure by successively broadening the period band from 100 - 200~s to 60 - 200~s. For this, we simulate 3-D viscoelastic wavefields on global fourth-order cubed-sphere-like meshes using the spectral-element solver Salvus \cite{afanasiev2019salvus} and reduce waveform fit iteratively using a mini-batch stochastic trust-region L-BFGS method \cite{vanherwaarden2020minibatch}. We quantify waveform misfits and compute adjoint sources using the time-frequency phase misfit \cite{fichtner2008time}. This allows us to eliminate the effect of unknown earthquake magnitude, and to incorporate both high-amplitude surface and low-amplitude body waves.\\[5pt]
In the following, we provide a detailed report on the construction of RevealLO, including data selection parameters and azimuthal coverage metrics (Section~\ref{sec:data}), forward solver implementation, regularization, and mesh reprojection (Section~\ref{sec:methodology}), and the resulting structural updates, comparisons with REVEAL, and waveform misfit validations (Section~\ref{sec:results}).


\section{Data Sources and Input Parameters}
\label{sec:data}

The construction of RevealLO relies on a global dataset of long-duration seismograms selected to capture multi-orbit surface waves and late body wave phases. Careful data curation is required when expanding the inversion window to 3.5~hour to ensure that the time series do not contain overlapping events or instrumental errors that may not be visible in 1~hour seismograms.


\subsection{Event Selection and Source Parameters}

To characterize the seismic sources, we retrieved Centroid Moment Tensor (CMT) solutions from the Global Centroid-Moment-Tensor (GCMT) catalog \cite{ekstrom2012global}. We selected 1235 unique earthquake events that occurred between 1982 and 2026, distributed globally to provide broad ray path coverage across the southern hemisphere and ocean basins.\\[5pt]
We selected events with magnitudes ($M_w$) between 6.5 and 7.5, to balance the signal-to-noise ratio (SNR) against source complexity for 3.5~hour records: For $M_w \le 6.5$, major-arc ($R_2, G_2$) surface wave signals are often below the ambient noise levels, and for $M_w \ge 7.5$, finite-source complexities increasingly conflict with the point-source approximation at 60 - 200~s.


\subsection{Azimuthal Path Diversity}
\label{sec:azimuthal_uniformity}

In seismic tomography, the lateral resolution and mitigation of trade-offs between isotropic wave speeds and unmodeled azimuthal anisotropy depend on the angular diversity of the crossing ray paths \cite{bezada2016anisotropy}. To quantify the improvement in surface wave coverage provided by extending the duration from 1~h to 3.5-h, this section introduces an azimuthal uniformity index.\\[5pt] 
Although body waves in a 1-h record already sample epicentral distances above $180^\circ$, fundamental-mode surface waves with group velocities $U_g \approx 3.8$~km/s are restricted to epicentral distances below $123^\circ$. Therefore, the analysis below specifically evaluates the azimuthal surface wave coverage gained by recording major-arc ($R_2, G_2$) surface wave arrivals in the 3.5-h window. We evaluate the azimuthal illumination potential of both durations, 1 h and 3.5 h. An event–station path is admissible if the complete dispersive surface wave train ($U_g \approx 3.8 -  4.4$~km/s) arrives within the respective duration at epicentral distances $\ge 15^\circ$. The calculation is performed across all accepted surface-wave windows through five steps, which are detailed in the following paragraphs.


\subsubsection{Directional Folding and Azimuthal Binning}

\begin{sidewaysfigure}
    \vspace{8.5cm}
    \centering
    \begin{tikzpicture}[scale=4]
        \hspace{-12cm}
        \includegraphics[width=0.8\textwidth]{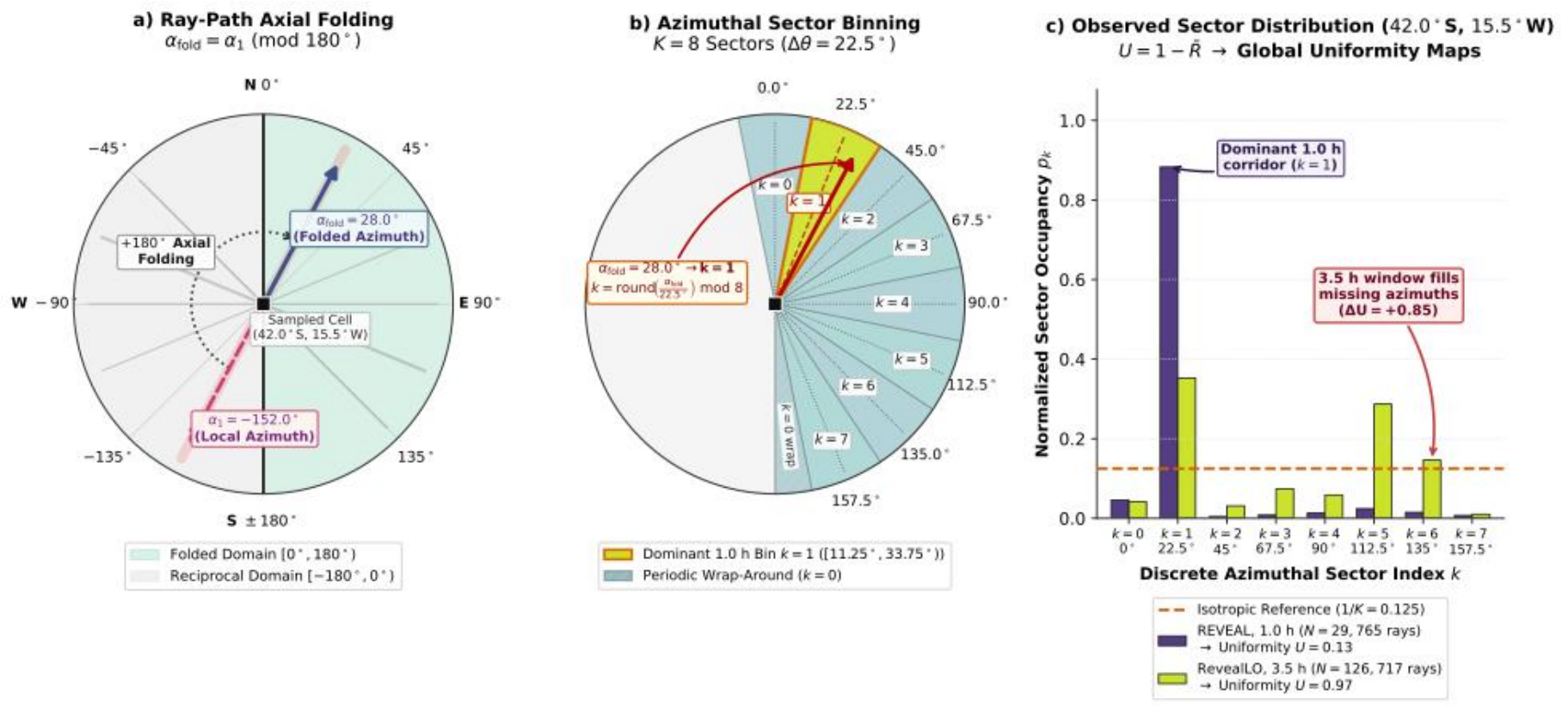} 
    \end{tikzpicture}
    \hspace{-5.5cm}
\caption{\textbf{Illustration of ray-path axial folding, discrete sector binning, and azimuthal uniformity evaluation for a single grid cell at $42.0^\circ S$~$15.5^\circ W$ $(\Delta~longitude, \Delta ~latitude)$.} 
\textbf{(a)} Axial folding maps an incoming ray path from the reciprocal domain ($\alpha_1 = -152^\circ$) into the folded domain ($\alpha_{\text{fold}} = 28^\circ \in [0^\circ, 180^\circ)$) using $\alpha_{\text{fold}} = \alpha_1 \pmod{180^\circ}$. 
\textbf{(b)} Discretization of the folded domain into $K = 8$ sectors ($\Delta\theta = 22.5^\circ$), showing the assignment of $\alpha_{\text{fold}} = 28^\circ$ to sector $k = 1$ and the periodic boundary condition for $k = 0$. 
\textbf{(c)} Resulting normalized sector occupancy $p_k$ comparing a corridor-dominated 1\,hour \textsc{REVEAL} distribution ($U = 0.13$, purple) against a near-uniform, omnidirectional 3.5\,hour \textsc{RevealLO} distribution ($U = 0.97$, green) relative to the ideal uniform threshold ($1/K = 0.125$, dashed line).}
\label{fig:azimuthal_binning}
\end{sidewaysfigure}

The seismic sensitivity to the velocity structure possesses axial symmetry -- where a wave propagating along the azimuth $\alpha$ provides the same constraints as one traveling along $\alpha + 180^\circ$ -- the propagation trajectories are evaluated modulo $180^\circ$ (Figure~\ref{fig:azimuthal_binning}a). Hence, for an earthquake-receiver pair with local path azimuth $\alpha_1 \in [-180^\circ, 180^\circ]$, the folded azimuth is defined as:
\begin{equation}
 \alpha_{\text{}} = \alpha_1 \pmod{180^\circ}\,.
\end{equation}
The continuous half-circle $[0^\circ, 180^\circ)$ is partitioned into $K = 8$ discrete azimuthal sectors of width $\Delta \theta = 22.5^\circ$, centered at discrete angles (Figure~\ref{fig:azimuthal_binning}b):
\begin{equation}
 \theta_k = k \cdot \frac{180^\circ}{K-1}, \quad k \in \{0, 1, \dots, K-1\}\,.
\end{equation}
Each accepted path segment is assigned to its nearest discrete sector index (Figure~\ref{fig:azimuthal_binning}c),
\begin{equation}
 k = \left\lfloor \frac{\alpha_{\text{}}}{180^\circ / K} + 0.5 \right\rfloor \pmod K\,.
\end{equation}


\subsubsection{Great-Circle Discretization: Minor vs.\ Major Arcs}

For each sector $k$, we map the surface wave trajectories onto a global geographic grid ($\Delta ~latitude = \Delta~longitude = 1.0^\circ$) using the following spherical great-circle binning:
\begin{enumerate}
 \item \textbf{1.0 h window (minor-arc $R_1 / G_1$ arrivals):} Paths are traced along the direct minor-arc great-circle geodesic between source and receiver up to a maximum surface-wave propagation distance  that corresponds to 1~h duration ($d \le 13\,680$~km, $\Delta \le 123^\circ$).
 
 \item \textbf{3.5 h window (minor-arc $R_1 / G_1$ + major-arc $R_2 / G_2$ arrivals):} In addition to minor-arc arrivals ($\Delta \le 180^\circ$), the window incorporates major-arc ($R_2, G_2$) surface waves. The major arc leaves the source at reverse azimuth $\alpha_{\text{maj}} = (\alpha_1 \pm 180^\circ) \pmod{180^\circ}$ and traverses the antipode over a great-circle distance $d_{\text{maj}} = 40\,075\text{ km} - d$.
\end{enumerate}
For each geographic cell $(i, j)$, accumulating the intersecting paths across all sectors, yields the sector counts $c_k(i, j)$, with the total path count $N(i, j) = \sum_{k=0}^{K-1} c_k(i, j)$.


\subsubsection{Azimuthal uniformity index}

To compute the directional dispersion of axial orientations without sign cancellation, we map angles to the full $360^\circ$ circle by angle doubling ($2\theta_k$). For every geographic cell $(i, j)$, the Cartesian components of the resulting directional vector are accumulated across the $K$ sectors:
\begin{equation}
 C(i, j) = \sum_{k=0}^{K-1} c_k(i, j) \cos(2\theta_k), \quad 
 S(i, j) = \sum_{k=0}^{K-1} c_k(i, j) \sin(2\theta_k).
\end{equation}
Because a cell with only two orthogonal paths yields a circular variance of 1 without providing statistically robust azimuthal sampling, we enforce a minimum ray-density threshold of $N(i, j) \ge N_{\min}$ (with $N_{\min} = 10$). For cells that meet this threshold, the mean resultant length $\bar{R} \in [0, 1]$ is
\begin{equation}
 \bar{R}(i, j) = \frac{\sqrt{C(i, j)^2 + S(i, j)^2}}{N(i, j)}\,.
\end{equation}
The quantity $\bar{R}$ measures the directional alignment. $\bar{R} = 1$ means that the paths that traverse the cell belong to a single azimuthal sector (unidirectional / parallel). In contrast, $\bar{R} = 0$ corresponds to trajectories that are distributed uniformly across azimuthal sectors. Based on this, the azimuthal uniformity index $U(i, j)$ is defined as the circular variance of the doubled angles:
\begin{equation}
 U(i, j) = 1 - \bar{R}(i, j) = 1 - \frac{\sqrt{\left[\sum_{k=0}^{K-1} c_k \cos(2\theta_k)\right]^2 + \left[\sum_{k=0}^{K-1} c_k \sin(2\theta_k)\right]^2}}{\sum_{k=0}^{K-1} c_k}\,.
 \label{eq:uniformity_index}
\end{equation}
Cells where $N(i, j) < N_{\min}$ are masked from the analysis. We then quantify the change in azimuthal uniformity introduced by assimilating 3.5 h instead of 1 h waveforms by $\Delta U(i, j)$
\begin{equation}
 \Delta U(i, j) = 
 \begin{cases}
  U_{3.5\text{}}(i, j) - U_{1.0\text{}}(i, j), & \text{if } N_{1.0\text{}} \ge N_{\min} \text{ and } N_{3.5\text{}} \ge N_{\min}, \\
  U_{3.5\text{}}(i, j), & \text{if } N_{1.0\text{}} < N_{\min} \text{ and } N_{3.5\text{}} \ge N_{\min}, \\
  \text{undefined (masked)}, & \text{otherwise}\,.
 \end{cases}
\end{equation}


\subsubsection{Tomographic Significance}

\begin{figure}[th!]
 \centering
 \includegraphics[width=1\textwidth]{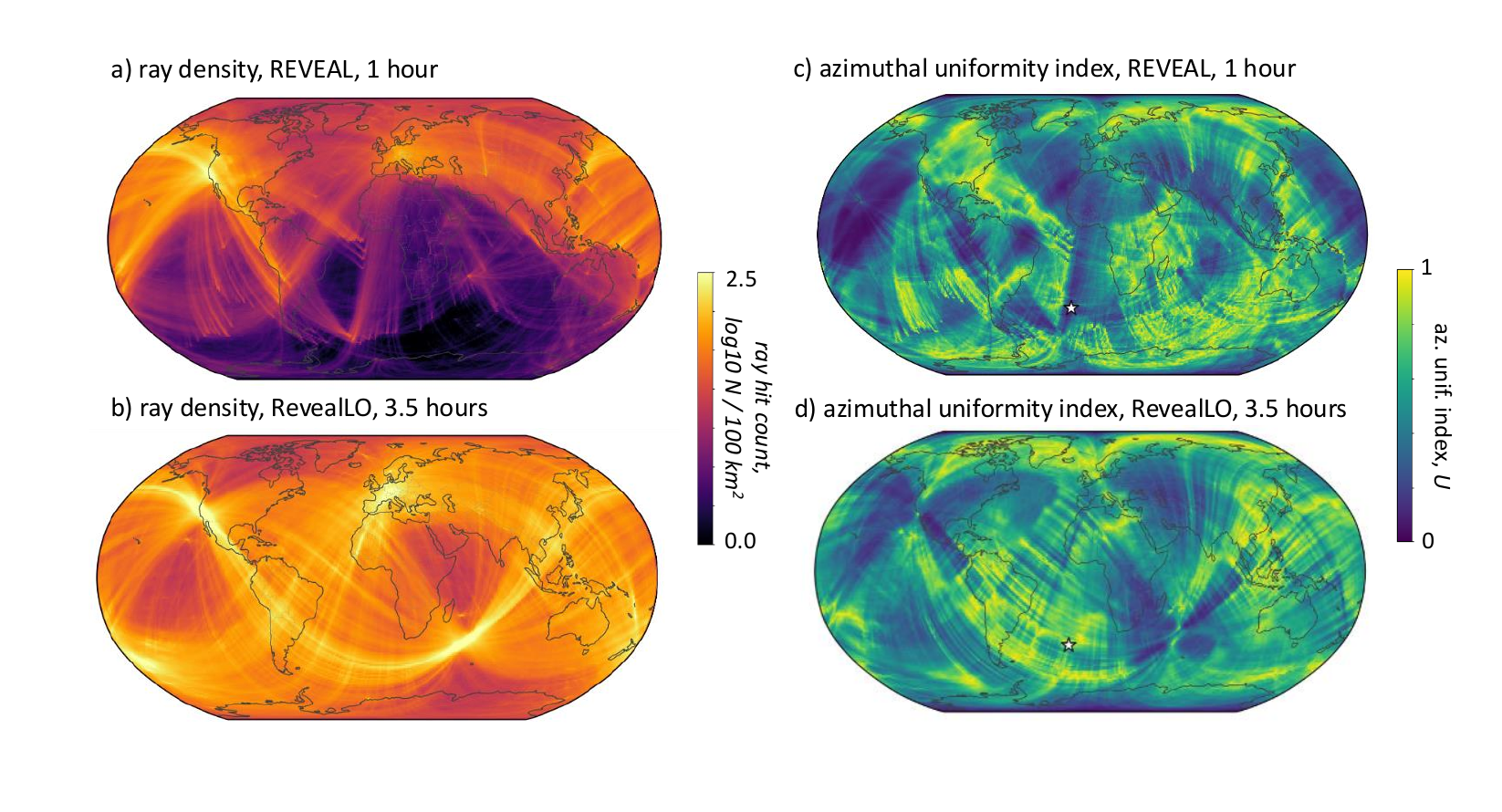}
 \caption{\textbf{Surface-wave ray density and azimuthal coverage for 1.0~h vs.\ 3.5~h recording windows.} 
  \textbf{(a, b)} Ray path density ($\log_{10} N / 100~km^2$) for baseline REVEAL (1.0~h, minor arcs only) and RevealLO (3.5~h, minor and major arcs). 
  \textbf{(c, d)} Azimuthal uniformity index ($U = 1 - \bar{R}$; 0 = unidirectional, 1 = isotropic) evaluated for cells with $N \ge 10$ paths. 
  The inclusion of major-arc $R_1$ orbital wave trains in RevealLO dramatically fills coverage gaps and elevates azimuthal uniformity across the Southern Ocean and oceanic basins. The white star marks the single cell used for the schematic in Fig.\ref{fig:azimuthal_binning}.}
 \label{fig:ray_azimuthal_uniformity}
\end{figure}

As shown in Fig.~\ref{fig:ray_azimuthal_uniformity}, the 1-hour surface-wave records provide high azimuthal uniformity primarily over densely instrumented continental regions in the northern hemisphere. In contrast, large areas of the South Pacific, Indian, and Southern Oceans exhibit low uniformity, dominated by sub-parallel ray corridors.\\[5pt]
The incorporation of additional minor- and major-arc surface waves in the 3.5 h seismograms yields an azimuthal information gain of $\Delta U \ge 0.5\text{--}0.7$ beneath these ocean basins. Since RevealLO is parameterized as a radially anisotropic medium (in terms of wave speeds $V_{\text{SV}}, V_{\text{SH}}, V_{\text{PV}}, V_{\text{PH}}$, and density $\rho$) and does not account for azimuthal anisotropy, unidirectional path coverage can map unmodeled azimuthal anisotropy to spurious isotropic velocity anomalies \cite{bezada2016anisotropy}. By supplying antipodal crossing paths across ocean basins, the 3.5 h data set averages directional biases and improves lateral resolution in data-sparse regions.


\subsection{Waveform Collection and Window Picking}
\label{S:windows}

\begin{sidewaysfigure}
    \vspace{8.5cm}
    \centering
    \begin{tikzpicture}[scale=4]
        \hspace{-13cm}
        \includegraphics[width=1.1\textwidth]{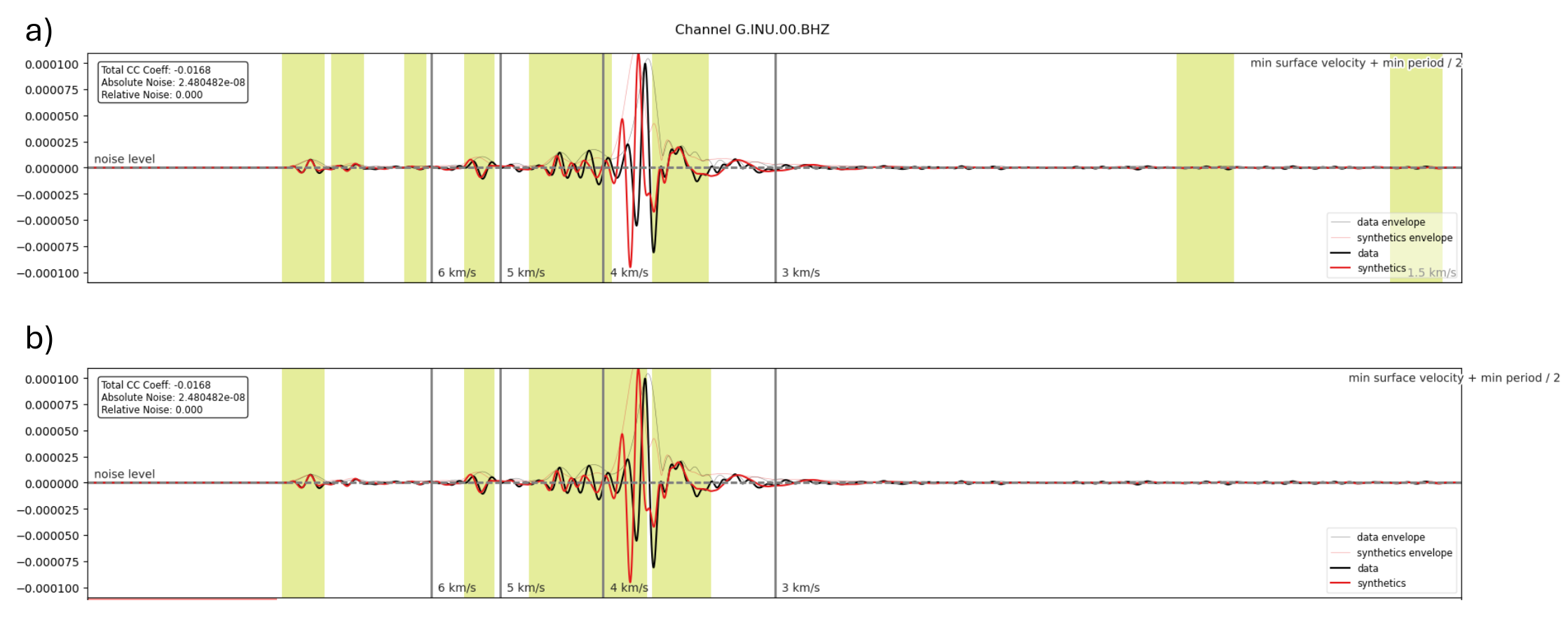} 
    \end{tikzpicture}
    \hspace{-2.5cm}
    \caption{\textbf{Illustration of automated window selection over extended 3.5 hour seismograms using LASIF 2.0.} The example is for a representative vertical-component record (\texttt{Channel G.INU.00.BHZ}). It compares the result of the algorithm for the settings used in the construction of REVEAL, and new settings adapted to long-period/long-duration recordings.
 \textbf{(a)} Standard windowing, as used for REVEAL, with conservative energy thresholds. The top trace in each panel displays observed displacement (black), computed displacement (red), envelopes (dashed), and accepted time windows (green shading).
 \textbf{(b)} Windowing as used for RevealLO, with a 10\% relative amplitude threshold, phase-dependent window lengths, and a maximum amplitude-to-window ratio. Time-local elimination criteria include cross-correlation coherence, maximum time shifts, envelope amplitude similarity, and energy thresholds. 
 Panel (b) additionally shows that standard thresholds reject spurious arrivals via the low energy criterion (bottom bar), isolating only body waves and the dominant direct surface wave.}
    \label{fig:lasif_windowing}
\end{sidewaysfigure}

For the 1235 selected events, we downloaded continuous three-component waveforms via the International Federation of Digital Seismograph Networks (FDSN) \cite{RomanowiczDziewonski1986}, yielding approximately 1.5 million unique ray paths. We corrected all waveform data for their instrument response, demeaned, detrended, and integrated to displacement prior to bandpass filtering.\\[5pt]
Prior to inversion, high-quality waveforms that match their corresponding computed waveforms within less than half a period need to be selected via an automated windowing procedure. For this, we used the window selection algorithm of the Large-scale Seismic Inversion Framework (LASIF) 2.0 \cite{thrastarson2022lasif}. It quantifies the similarity between observed and computed waveforms to define time windows suitable for time-frequency phase misfit evaluation. The selection process operates in two stages: global trace acceptance and localized window elimination. An example of the procedure is visualized in Fig.~\ref{fig:lasif_windowing}.
%
%


\subsubsection{Global Trace Acceptance}

Before individual phases are windowed, the full 3.5 h seismogram must pass basic quality control. A trace is rejected if it fails either of the following criteria:
\begin{itemize}
 \item \textbf{Global Cross-Correlation:} The normalized cross-correlation coefficient between the complete observed and computed traces must exceed 0.10.
 \item \textbf{Relative Noise Level:} The maximum pre-event noise level (measured from absolute amplitudes prior to the first P-wave arrival) must not exceed 10\% of the maximum amplitude of the full trace.
\end{itemize}


\subsubsection{Local Window Selection Criteria}

For traces that pass the global acceptance criteria, the algorithm evaluates candidate windows between local envelope minima, eliminating segments that may exhibit cycle skipping, low SNR or dissimilar amplitudes. The complete suite of criteria is listed below. The choice of specific numbers is the result of careful empirical tests.
%
\begin{itemize}
 \item \textbf{Phase Coherence (Correlation):} The local normalized cross-correlation coefficient within a candidate window must exceed 0.65.
 \item \textbf{Time-Shift Constraints:} To prevent cycle skipping, the maximum allowable lag time at peak cross-correlation between observed and computed phases must not exceed 35\% of the minimum period.
 \item \textbf{Envelope Similarity:} A minimum envelope amplitude ratio of 0.2 is enforced, to avoid accidental fitting of low-amplitude noise.
 \item \textbf{Energy and Peak Amplitude Ratios:} The maximum allowable energy ratio between observed and computed waveforms within a time window is capped at 10. Furthermore, to retain low-amplitude body waves and major-arc arrivals while rejecting background noise, a candidate window must have a peak amplitude of at least 10\% of the overall trace maximum amplitude.
 \item \textbf{Peak and Trough Matching:} Windows are dynamically separated or merged using a Peak and Trough matching algorithm that associates local extrema. It prevents adjacent phases with different dispersion characteristics from merged into a single window.
 \item \textbf{Phase-Dependent Window Lengths:} Minimum window durations are scaled by group velocity. Surface waves are assigned a minimum window length of $2.4\,T_{\min}$ to capture dispersive wave trains, especially of major-arc $R_2$ and $G_2$ waves.
\end{itemize}

\section{Model Construction and Methodology}
\label{sec:methodology}

The RevealLO inversion framework follows the multi-scale, full-waveform inversion workflow established for REVEAL \cite{thrastarson2024}. We use the spectral-element solver Salvus \cite{afanasiev2019salvus} for 3-D viscoelastic wave propagation, the time-frequency phase misfit \cite{fichtner2008time} for the quantification of waveform differences and adjoint source calculation, and the L-BFGS algorithm of the mini-batch stochastic trust-region \cite{vanherwaarden2020minibatch} for iterative optimization.

\subsection{Model Parameterization and Optimization Schedule}
\label{sec:parameterization}

The Earth model is parameterized on a fourth-order cubed-sphere mesh, capable of resolving a minimum period of 60~s. The inversion updates four radially anisotropic elastic wave speed parameters: vertically and horizontally polarized S wave velocities ($V_{\text{SV}}, V_{\text{SH}}$) and P wave velocities ($V_{\text{PV}}, V_{\text{PH}}$), while density ($\rho$) and attenuation ($Q_\mu, Q_\kappa$) are kept fixed to the starting model, i.e., to the values of REVEAL.\\[5pt]
To mitigate non-linearity and prevent cycle-skipping, the optimization was performed sequentially across five period bands, starting at 100--200~s and ending at 60--200~s. The total number of iterations, summarized in Table~\ref{tab:inversion_bands}, is 135.\\[5pt]
%
\begin{table}[t]
  \centering
  \small
  \caption{\textbf{Sequential multi-scale inversion schedule for RevealLO.} The iterative optimization was based on a mini-batch stochastic trust-region L-BFGS algorithm on a tensor-order 4 cubed-sphere mesh.}
  \label{tab:inversion_bands}
  \vspace{0.3cm}
  \begin{tabular}{lccc}
    \toprule
    \textbf{Inversion Stage} & \textbf{Period Band (s)} & \textbf{Iterations} & \textbf{Misfit Reduction (\%)} \\
    \midrule
    Band 1 & 100--200 & 70 & 12.0\% \\
    Band 2 &  90--200 & 30 &  2.0\% \\
    Band 3 &  80--200 & 20 &  1.0\% \\
    Band 4 &  70--200 & 10 &  0.5\% \\
    Band 5 &  60--200 &  5 &  0.1\% \\
    \bottomrule
  \end{tabular}
\end{table}
%
The optimization started with 70 iterations in the 100 - 200 s period band. It used quasi-random mini-batches of events, retaining an overlapping subset between consecutive iterations to monitor stochastic gradient consistency \cite{vanherwaarden2020minibatch}. This initial long-period stage established the primary large-scale structural corrections, achieving a waveform misfit reduction of 12.0 \%. Subsequently, we broadened the period band and iterated until the misfit curve flattened. The misfit reduction for each band decreased successively, reaching 0.1 \% in the 60 - 200 s band. We stop at 60 s, because for shorter periods major-arc surface waves tend to have insufficiently large amplitudes at epicentral distances $>270^\circ$.\\[5pt]
Before each model update, raw sensitivity kernels were smoothed using an anisotropic diffusion operator with a lateral smoothing length of $0.5 \times \lambda_{\min}$ and a vertical smoothing length of $1.0 \times \lambda_{\min}$, where $\lambda_{\min}$ is the local minimum wavelength of S waves in a given period band.

\subsection{Initial Model Preparation}
\label{sec:baseline_prep}
\begin{figure}[th!]
 \centering
 \includegraphics[width=0.85\textwidth]{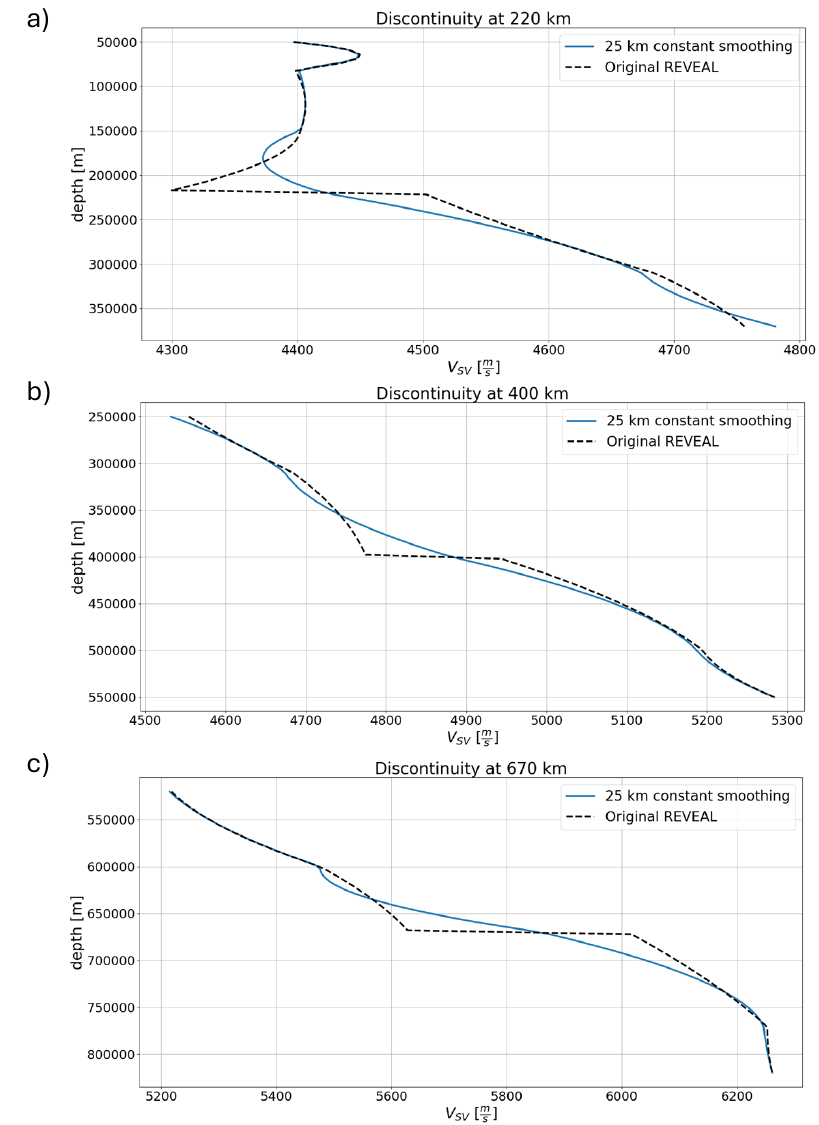}
 \caption{ \textbf{Mantle discontinuity smoothing applied to REVEAL.} 
 Vertical profiles of $V_{\text{SV}}$, comparing the original REVEAL (blue solid) against the smoothed version used as initial model for RevealLO (black dashed). 
 \textbf{(a)} the 220~km Lehmann discontinuity, 
 \textbf{(b)} the 400~km olivine-to-wadsleyite transition, and 
 \textbf{(c)} the 670~km ringwoodite-to-bridgmanite transition. }
 \label{fig:discontinuity_smoothing}
\end{figure}

\begin{figure}[th!]
 \centering
 \includegraphics[width=0.95\textwidth]{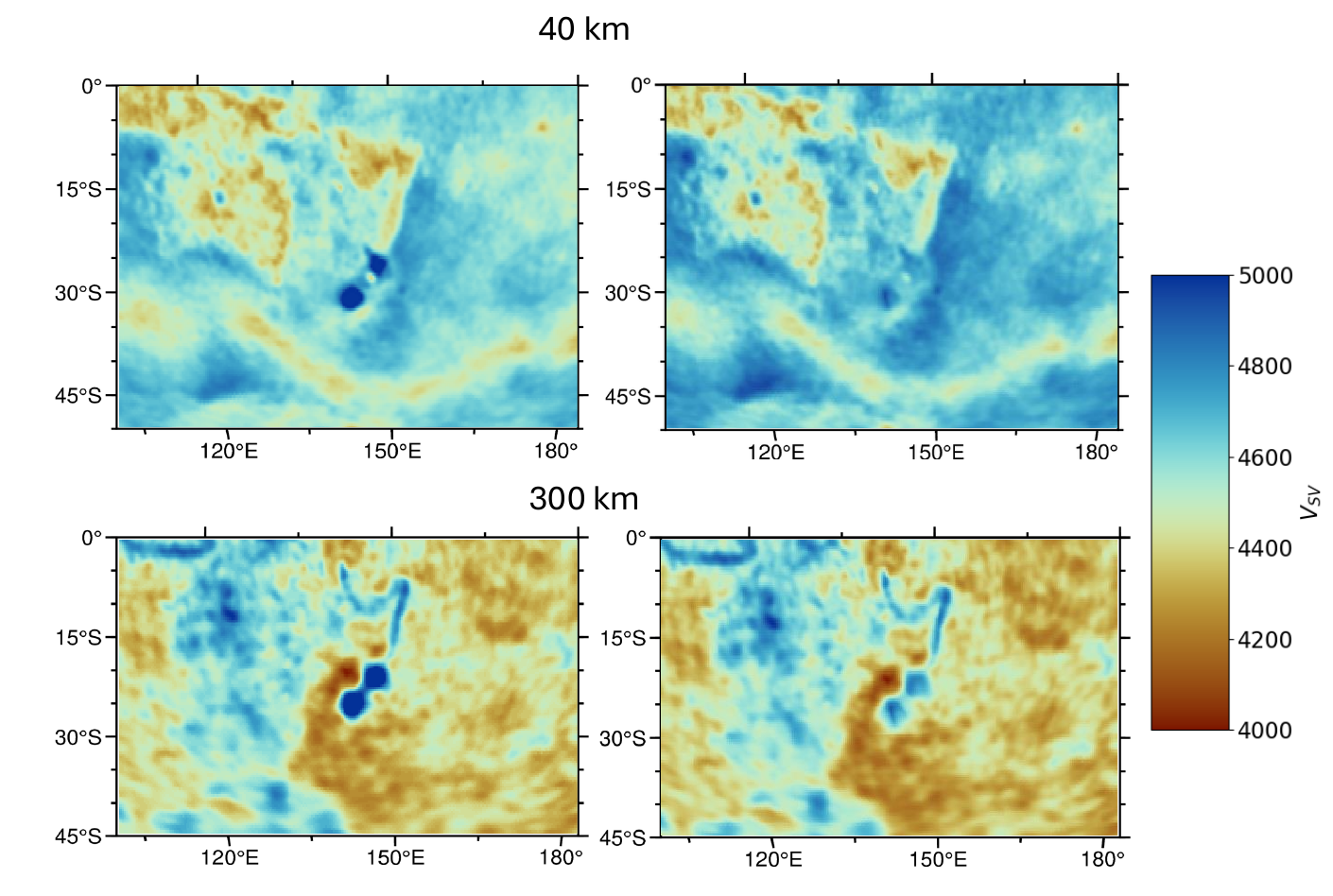}
 \caption{\textbf{Gaussian tapering for New Zealand artifact mitigation in REVEAL.} Regional shear velocity maps before (left) and after (right) applying the Gaussian taper at 40 km (top) and 300 km (bottom) depth.}
 \label{fig:nz_artifact}
\end{figure}

Despite its ability to explain global wave propagation at 33 s period with high accuracy, we opted to precondition two features in the starting model to let the extended 3.5h dataset determine their structure. The two most important are mantle discontinuities inherited from PREM \cite{prem1981} and a high wave speed artifact beneath New Zealand.
\paragraph{Mantle Discontinuity Smoothing:} REVEAL was constructed on the basis of the long-wavelength full-waveform model LOWE \cite{thrastarson2022}, which in turn used PREM \cite{prem1981} as the initial model. Consequently, REVEAL inherited mantle discontinuities at 220 km, 400 km and 670 km depth. Since these discontinuities are externally imposed instead of being constrained by our dataset, we decided to remove them to gain data-determined discontinuities. For this, we apply a vertical smoothing filter with 25 km cutoff that replaces them with a gradual transition. The result is visualized in Fig.~\ref{fig:discontinuity_smoothing}. The removal of the discontinuities enables the data to produce velocity gradients that are as sharp as required, and possibly located at different depths than previously imposed.
%
%
\paragraph{New Zealand Artifact Mitigation:} The REVEAL inspection identified a localized, high-amplitude velocity anomaly around New Zealand that extends through the upper mantle (Fig.~\ref{fig:nz_artifact}, left). Comparisons with regional tomographic models suggest that this characteristic is an inversion artifact. It was likely acquired by fitting erroneous ocean-bottom seismometer data during the final short-period inversion stages of REVEAL.\\[5pt]
To remove this feature, we applied a localized, depth-dependent taper, constructed from three 3-D Gaussian functions centered beneath New Zealand at depths of 100~km (400~km half-width), 350~km (150~km half-width) and 500~km (300~km half-width). The application of the taper reduced the velocities back to the original background mean (Fig.~\ref{fig:nz_artifact}, right). During the construction of RevealLO, no additional tapering was applied to this region, allowing the long-period surface and body waves to reconstruct the data-consistent regional structure.
%
%

\subsection{Increased Depth Sensitivity}
\label{sec:adjoint_gradients}

\begin{figure}[th!]
 \centering
 \includegraphics[width=1\textwidth]{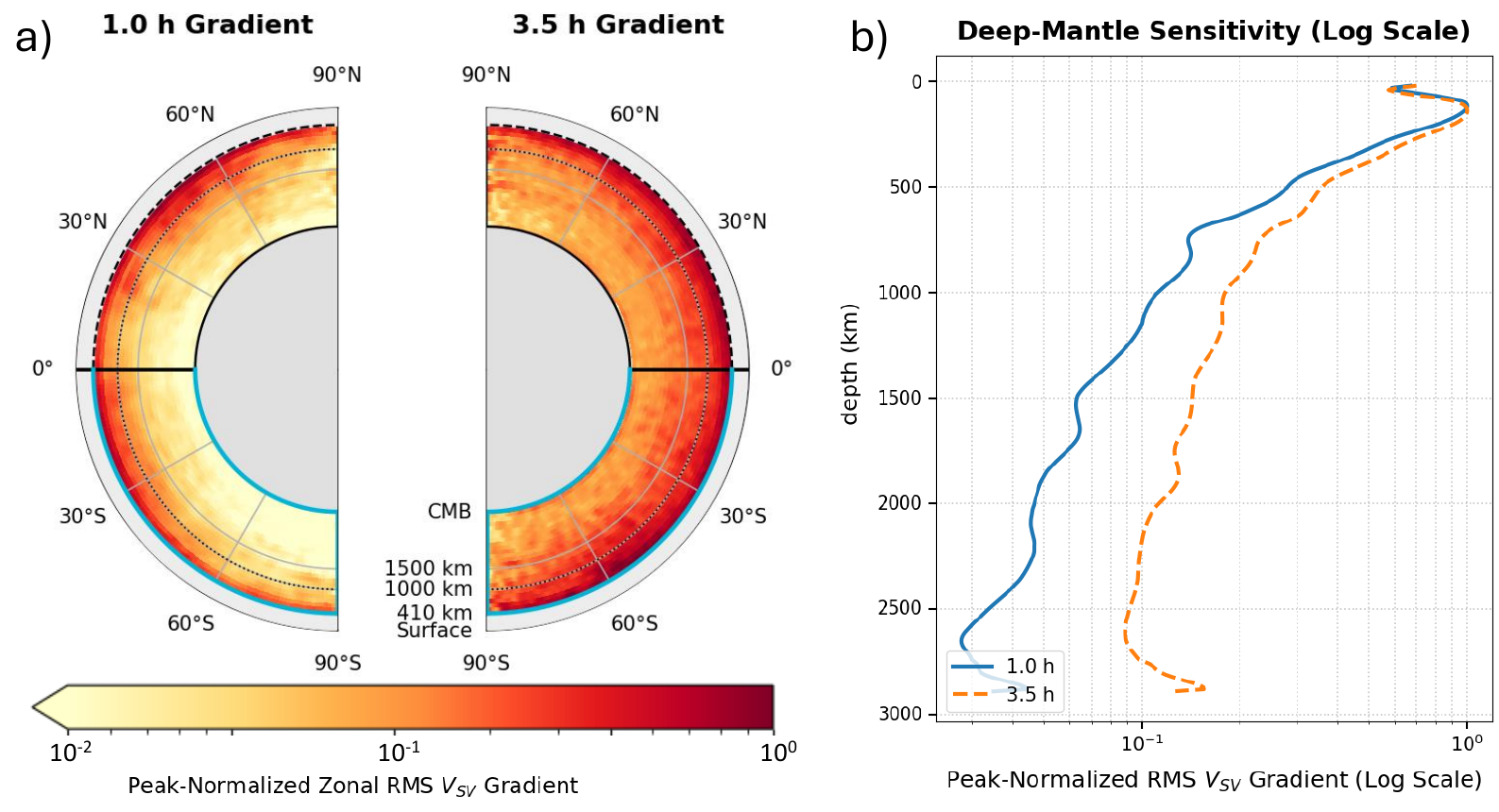}
 \caption{\textbf{Comparison of the cumulative sensitivity distribution for 80 randomly selected events based on 1 h and 3.5 h recordings.}  \textbf{(a)} Longitudinally averaged $V_\text{SV}$ sensitivity amplitudes (RMS) for 1 h (left) and 3.5 h (right) seismograms, normalized with respect to the peak amplitude. \textbf{(b)} Spherically averaged (depth-dependent) sensitivity amplitudes, normalized to the peak value and displayed logarithmically.}
 \label{fig:adjoint_gradients}
\end{figure}

Expanding the duration of the seismogram from 1 to 3.5 hours, alters the sensitivity distribution, as shown in the $V_{\mathrm{SV}}$ sensitivity examples in Fig.~\ref{fig:adjoint_gradients}. In an inversion with 1 h data, cumulative sensitivity of 80 randomly selected events is concentrated along the direct body wave and the minor-arc surface wave paths ($R_1, G_1$). By applying the LASIF~2.0 windowing strategy (Section~\ref{S:windows}) to 3.5 h records, the measurements extend to additional wavefield components that increase sensitivity at different depths.
\begin{enumerate}
 \item \textbf{Upper Mantle ($\sim$50 -- 400 km depth):} The incorporation of major-arc ($R_2, G_2$) surface waves provides additional paths in the hemisphere opposite the source. The effect is particularly pronounced in the southern hemisphere and within the upper 400 km, where the sensitivity to surface waves is the highest. Expected benefits are improved resolution of the upper-mantle structure and reduced trade-offs with unmodeled azimuthal anisotropy.
 \item \textbf{Transition Zone and Mid-Mantle ($\sim$400 -- 1500~km depth):} The 3.5 h window includes additional multiply reflected body wave arrivals. Because the time-frequency phase misfit normalizes amplitude differences, these lower-amplitude phases can contribute to the gradient, 
 provided that their peak envelope amplitude is $>10\%$ the trace maximum. In the horizontally averaged RMS gradient profiles of Fig.~\ref{fig:adjoint_gradients}b, this manifests itself as a pronounced increase in sensitivity across the mantle transition zone and the mid-mantle.
 \item \textbf{Lower-Mantle ($\gtrsim$1500~km depth):} Improvements in lower-mantle sensitivity also derive from the incorporation of late-arriving body wave phases. Sensitivity near the CMB, in particular, benefits from multiple core reflections, such as $ScS_n$.
\end{enumerate}
%
%

\subsection{Model Integration and Mesh Reprojection}
\label{sec:reprojection}

RevealLO inversion was conducted on a fourth-order cubed-sphere mesh designed for a minimum period of 60~s, the resulting long-period updates must be transferred to the high-resolution REVEAL baseline model, which is defined on a second-order mesh for a minimum period of 33~s. To preserve the short-wavelength (33 -- 60~s) structural details of REVEAL while incorporating the 60 -- 200~s RevealLO updates, we applied a four-step mesh reprojection procedure:
\begin{enumerate}
 \item \textbf{Forward Reprojection:} The REVEAL model $m_0$ was interpolated from its native  33~s mesh onto the coarser 60~s mesh, to produce an initial model $m'_0$ for RevealLO inversion.
 \item \textbf{Difference Calculation:} We isolated the total update generated by the 3.5 h inversion by computing the model difference on the RevealLO mesh, $\Delta m'_{\text{update}} = m'_{\text{inverted}} - m'_0$.
 \item \textbf{Adjoint/Interpolative Reprojection:} The isolated long-period update $\Delta m'_{\text{update}}$ was interpolated back onto the 33~s REVEAL mesh to obtain $\Delta m_{\text{update}}$.
 \item \textbf{Model Integration:} Following lowermost-mantle tapering (Section~\ref{sec:cmb_taper}), the reprojected update was added to REVEAL, to produce the final RevealLO model:
 \begin{equation}
  m_{\text{}}^{\text{}} = m^{\text{}}_0 + \Delta m^{\text{}}_{\text{update}}.
 \end{equation}
\end{enumerate}

\subsection{Deep Mantle Regularization and CMB Tapering}
\label{sec:cmb_taper}

\begin{sidewaysfigure}
    \vspace{12.5cm}
    \centering
    \begin{tikzpicture}[scale=4]
        \hspace{-13cm}
        \includegraphics[width=1.17\textwidth]{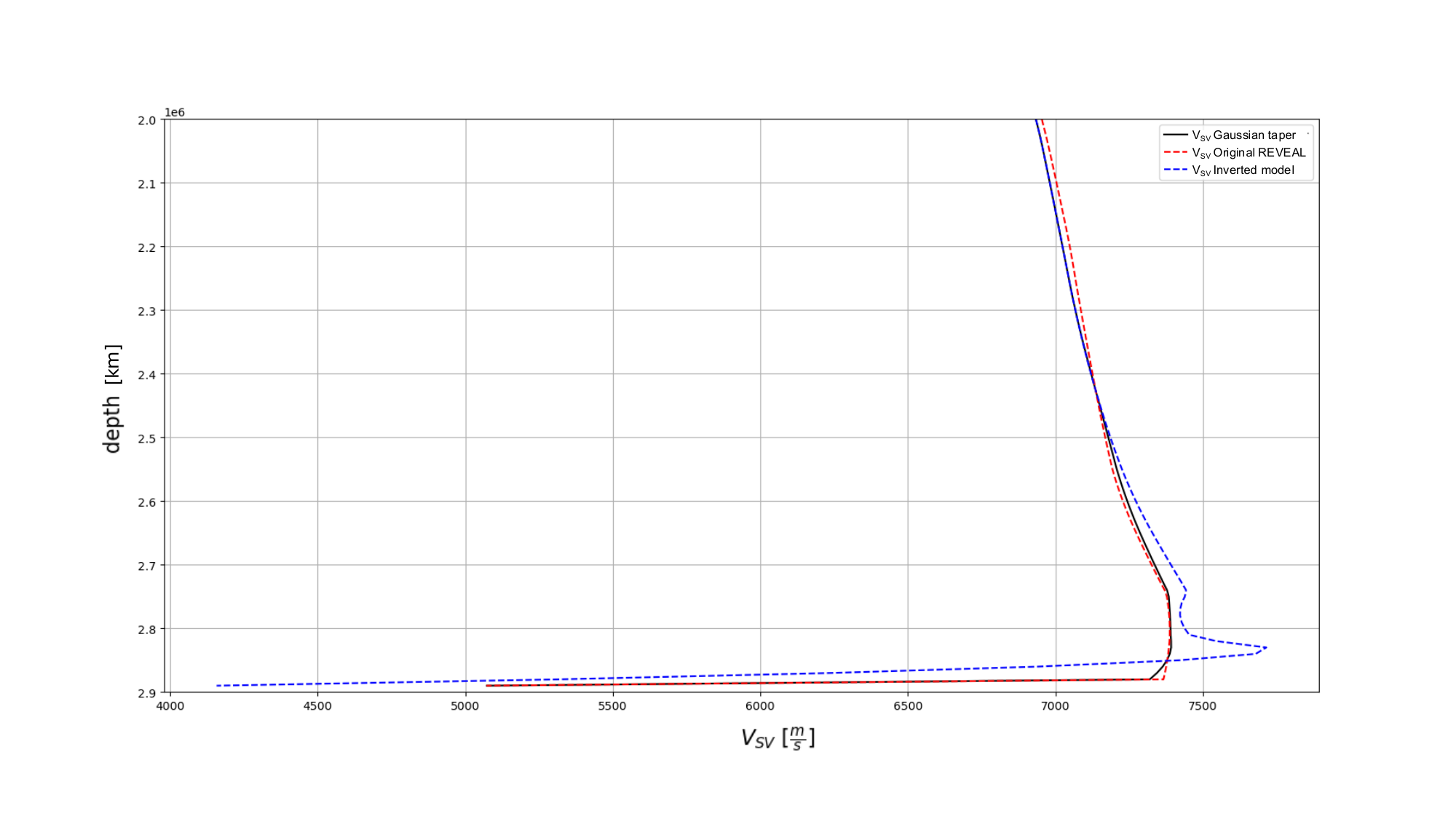} 
    \end{tikzpicture}
    \caption{\textbf{Removal of lowermost mantle artifact.} 
 Spherically averaged vertical profiles of $V_{\text{SV}}$ from 2000~km depth to the CMB. The raw inverted model (dashed blue line) exhibits unusual boundary oscillations near the CMB, featuring a high-velocity excursion followed by a steep drop. To remove this artifact, a depth-dependent Gaussian taper (solid black line) is applied below 2300~km depth, smoothly returning the model update to the baseline REVEAL wave speeds (dashed red line) at the CMB.}
    \label{fig:cmb_taper}
\end{sidewaysfigure}

Inspection of the raw inverted model revealed an unusual velocity oscillation located in the lower mantle, close to the CMB, shown in Fig.~\ref{fig:cmb_taper}. This artifact arises from the sparse long-period S wave sensitivity in the $D''$ region combined with smoothing of the vertical gradient across the solid-fluid CMB interface. To remove this technical construction error, we applied a depth-dependent Gaussian taper to the model update $\Delta m_\text{update}(z)$ between 2300~km depth and the CMB:
\begin{equation}
  \Delta m_{\text{update}}(z) \cdot \exp\left( -\frac{(z - 2300\,\text{})^2}{2\sigma_z^2} \right) \to \Delta m_{\text{update}}(z)\,, \quad z \ge 2300\text{~km},
 \label{eq:cmb_taper}
\end{equation}
where $\sigma_z = 350$~km. As shown by the solid black curve in Fig.~\ref{fig:cmb_taper}, the taper damps the boundary oscillation and returns the velocity structure below $\sim 2300$~km smoothly to the REVEAL.
%
%

\section{Results and Model Validation}
\label{sec:results}

The following sections summarize structural features of RevealLO relative to REVEAL, and the ability of RevealLO to explain long-period, long-duration recordings.

\subsection{Amplitude and Spatial Distribution of Structural Updates}
\label{sec:model_comparison}

Fig.~\ref{fig:reveal_diff} provides a comparison of RevealLO and REVEAL in terms of the Voigt-averaged isotropic S wave speed $V_S$ RMS changes as a function of depth. Comparing the RMS amplitude of $V_\text{S}$ variations in REVEAL and RevealLO (Fig.~\ref{fig:reveal_diff}a) and depth slices  (Fig.~\ref{fig:reveal_diff}b--f), reveals three different depth regimes:
%
\begin{sidewaysfigure}
    \vspace{12.5cm}
    \centering
    \begin{tikzpicture}[scale=4]
        \hspace{-10cm}
        \includegraphics[width=1\textwidth]{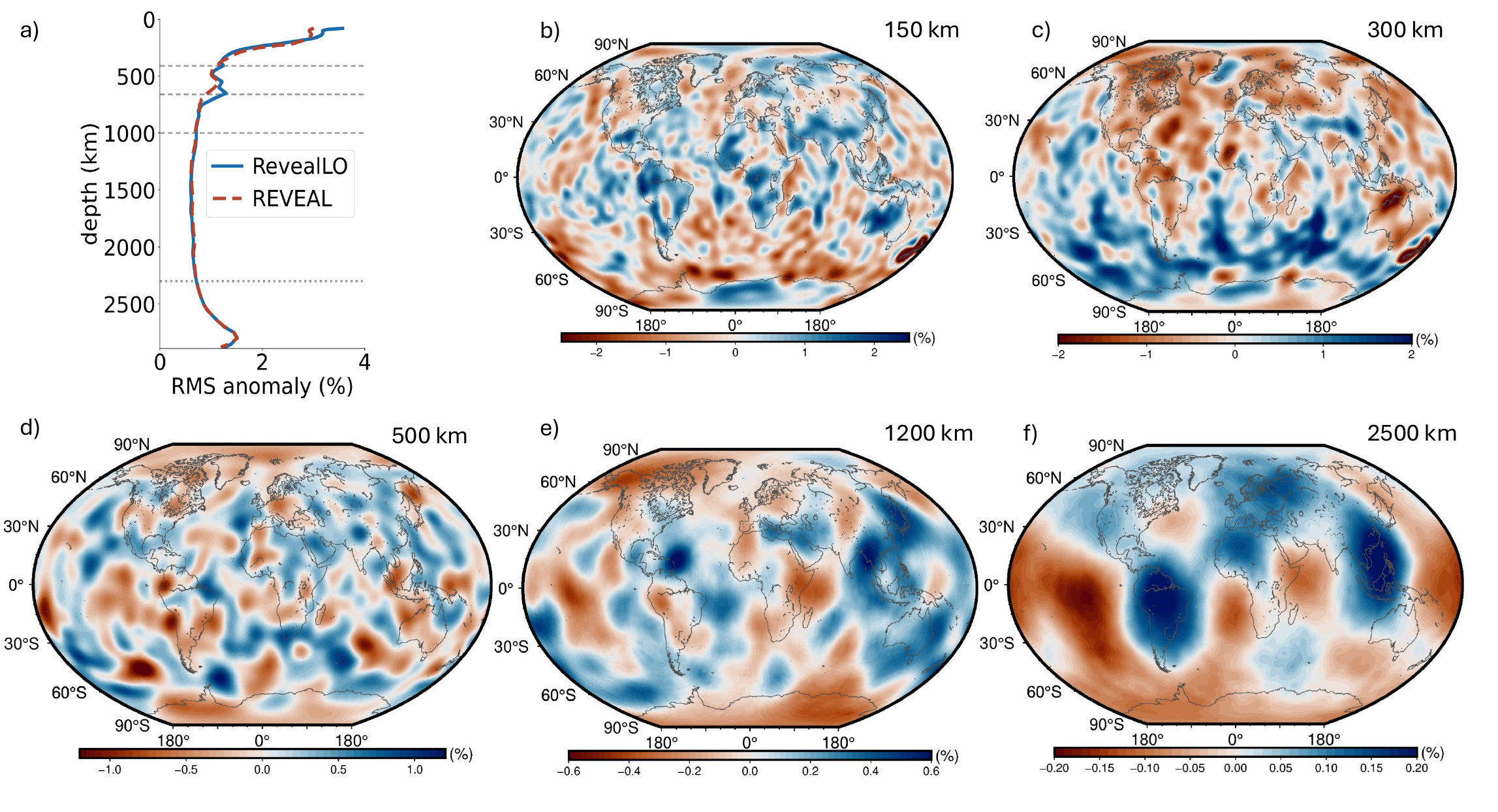} 
    \end{tikzpicture}
    \caption{\textbf{Depth-dependent comparison between RevealLO and REVEAL.} 
\textbf{(a)} The RMS anomaly amplitudes of REVEAL and RevealLO as a function of depth. \textbf{(b--f)} Global map views of the $dlnV_\text{S}$ difference between REVEAL and RevealLO at depths of 150~km, 300~km, 500~km, 1200~km and 2500~km}
    \label{fig:reveal_diff}
\end{sidewaysfigure}
%
\begin{itemize}
 \item \textbf{Upper Mantle and Transition Zone (100--700~km):} The structural updates are largest in this interval, where the RMS difference between RevealLO and REVEAL amounts to several tens of percent of REVEAL's RMS anomaly amplitude. The southern hemisphere experiences larger modifications than the northern hemisphere, especially within the 100--200~km depth interval, most likely due to the addition of major-arc surface wave paths across southern ocean basins.
 \item \textbf{Lower Transition Zone to Mid-Mantle (400--1500~km):} Below 400~km depth, the northern and southern hemispheres change by nearly equal amounts. Between 1000--1500~km depth, the RMS difference between RevealLO and REVEAL is $\sim$25 \% of REVEAL's anomaly amplitude. Because fundamental-mode surface waves have negligible sensitivity at these depths, this mid-mantle update is likely dominated by surface wave overtones and multiply reflected body-wave phases ($SS, SSS, ScS$ multiples) included in the 1--3.5~h time windows.
 \item \textbf{Lowermost Mantle ($> 2300$~km):} Below 2300~km depth, the model difference drops below 0.3 \% of REVEAL's RMS anomaly amplitude. This is where the Gaussian taper (Eq.~(\ref{eq:cmb_taper})) smoothly transitions RevealLO back into exact agreement with REVEAL at the CMB.
\end{itemize}

\subsection{Global Mantle Structure and Radial Anisotropy}
\label{sec:tectonic_features}
%
\begin{figure}[p!]
 \centering
 \includegraphics[width=0.95\textwidth]{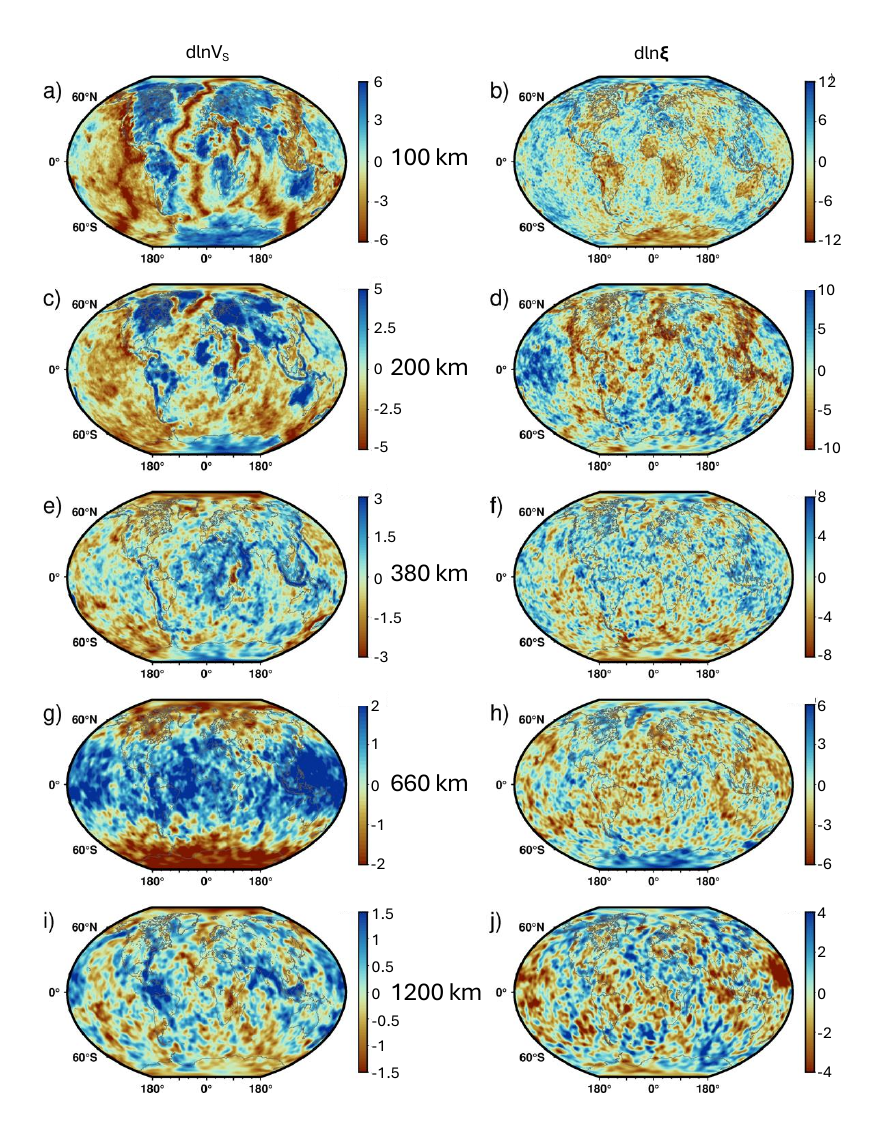}
 \caption{\textbf{Isotropic S-wave structure and radial anisotropy in RevealLO.} \textbf{Left column:} Voigt-averaged isotropic S-wave velocity perturbations ($\delta \ln V_\text{S}$) relative to 1-D depth-shell means at 100, 200, 380, 660 and 1200~km depth. \textbf{Right column:} Radial anisotropy variations ($\delta \ln \xi$) relative to depth-shell means. Blue indicates $V_{\text{SH}} > V_{\text{SV}}$, and red $V_{\text{SV}} > V_{\text{SH}}$.}
 \label{fig:depth_slices_reveal_lo}
\end{figure}
%
Fig.~\ref{fig:depth_slices_reveal_lo} displays the absolute Voigt-averaged isotropic S-wave velocity anomalies ($\delta \ln V_\text{S}$) along with radial anisotropy variations ($\delta \ln \xi$, where $\xi = V_{\text{SH}}^2 / V_{\text{SV}}^2$) in RevealLO at various depths. In the upper mantle (100--200~km), RevealLO delineates high-velocity cratonic lithosphere ($\delta \ln V_\text{S} > 4 \%$) and low-velocity anomalies along mid-ocean ridge systems and back-arc basins ($\delta \ln V_\text{S} < -4$ \%). Radial anisotropy between 100--200~km depth exhibits positive perturbations beneath the Pacific, Atlantic and Indian oceans with negative perturbations beneath continental shields and convergent margins. Across the transition zone (380--660~km), circum-Pacific subduction zones appear as continuous high-velocity features, while at 1200~km depth, the model resolves mid-mantle wave speed heterogeneity and moderate radial anisotropy ($\pm 2$ \%) above the Pacific and African large low-shear-velocity provinces (LLSVPs).

\subsection{Seismic Waveform Fit and Cross-Validation}
\label{sec:waveform_validation}

\begin{sidewaysfigure}
    \vspace{12.5cm}
    \centering
    \begin{tikzpicture}[scale=4]
        \hspace{-12cm}
        \includegraphics[width=1.1\textwidth]{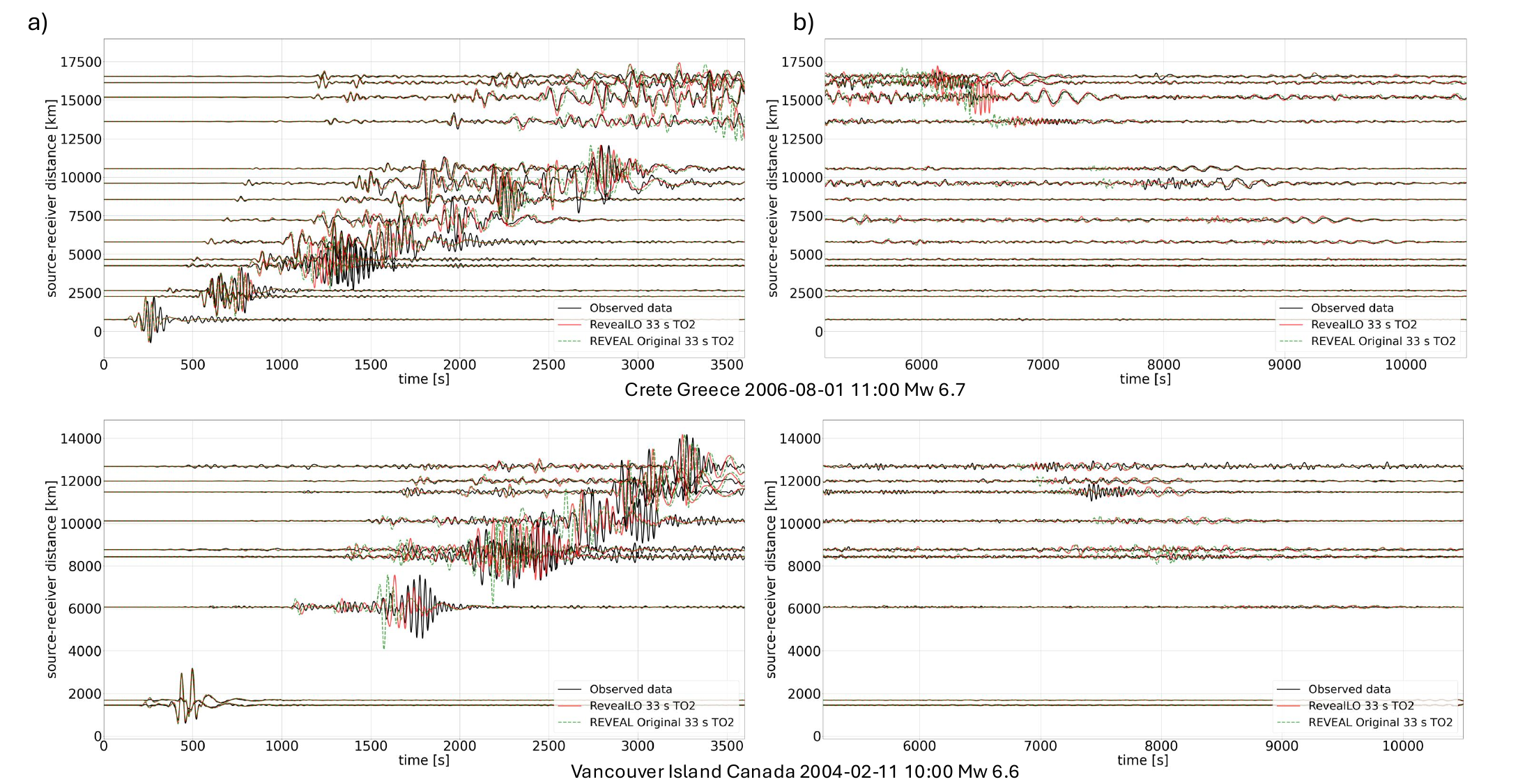} 
    \end{tikzpicture}
    \caption{\textbf{Comparison of selected observed and synthetic seismograms for the Crete ($M_w$~6.7) and Vancouver Island ($M_w$~6.6) earthquakes. } The time series show observed (black), synthetic REVEAL (dashed green) synthetic RevealLO (red) seismograms with a minimum period of 33~s. Panel \textbf{(a)} focuses on minor-arc waveform fits, while Panel \textbf{(b)} focuses on major-arc waveform fits for the two example events. RevealLO preserves the fit to early body wave arrivals while systematically correcting phase shifts and amplitude mismatches in regional and teleseismic surface wave trains ($t > 1500$~s). Meanwhile the major-arc segments show improvements on late surface wave arrivals for most epicentral distances.}
    \label{fig:waveform_fits}
\end{sidewaysfigure}

To quantify the ability of RevealLO to fit seismic data, we evaluated its waveform fit with two complementary tests:  (i) cross-validation against an independent control set of earthquakes not used in the inversion, and (ii) standard 1 h $L_2$ waveform misfit evaluation at short periods (33--200~s) before and after mesh reprojection.
\begin{enumerate}
 \item \textbf{Validation Data:} To verify that the model generalizes beyond the 1\,235 events used in the inversion, we calculated 3.5 h synthetic seismograms for a validation dataset of 54 independent earthquakes not included in the optimization. For this validation data set, RevealLO reduces the overall $L_2$ waveform misfit by 15.6 \% relative to the starting REVEAL model. 
 \item \textbf{Short-Period (33~s) 3.5 h Validation:} A key requirement of mesh reprojection (Section~\ref{sec:reprojection}) is that adding the 60--200~s updates from the long-period inversion mesh back onto the original REVEAL mesh must not degrade the 33~s waveform fit established by REVEAL. We performed forward simulations on the REVEAL mesh up to the 33~s period using 3.5 h seismograms. Evaluated across all 1 h body-wave and minor-arc windows, the reprojected RevealLO model yields an overall improved $L_2$ waveform misfit compared to the original REVEAL. This confirms that the reprojection strategy and CMB tapering preserve shorter-period phase alignments while integrating the long-period 3.5 hour updates. Fig.~\ref{fig:waveform_fits} shows example waveform fits comparing the waveform simulation ability of the original REVEAL model compared to the reprojected RevealLO at 33 s. RevealLO shows improvements in the fitting of surface wave overtones compared to the original REVEAL model.
\end{enumerate}
%
%

%

\section{Discussion and Limitations}
\label{sec:discussion}

Extending the duration of the inverted record from 1~h to 3.5~h in RevealLO increases the azimuthal coverage of surface waves in the southern hemisphere and the major ocean basins. By incorporating major-arc ($R_2, G_2$) surface waves along late-arriving body wave multiples, the update modifies isotropic shear-wave speeds by 30--70 \% of the anomaly amplitude of REVEAL in the upper mantle and transition zone, and by 25--33 \% in the mid-mantle (1000--1500~km). Several methodological limitations should be noted when interpreting RevealLO:
\begin{enumerate}
 \item \textbf{Depth Sensitivity Decay:} Because fundamental-mode surface wave kernels decay exponentially with depth, the enhanced azimuthal coverage of the southern hemisphere primarily benefits the upper 400~km of the mantle. Below the transition zone, structural updates rely on surface wave overtones and multiply reflected body waves, and below 2300~km the model is tapered back to REVEAL to suppress CMB boundary oscillation artifacts.
 \item \textbf{Radial Anisotropy Parameterization:} RevealLO is parameterized as a radially anisotropic medium and does not explicitly invert for azimuthal anisotropy parameters. Although the improved azimuthal distribution of the crossing ray paths in the 3.5 h dataset mitigates the mapping of unmodeled azimuthal anisotropy into spurious isotropic anomalies \cite{bezada2016anisotropy}, full recovery of the 3-D anisotropic structure will require explicit azimuthal parameterization in future updates.
 \item \textbf{Minimum Period of the Extended Update:} The 3.5 h inversion was performed at periods of 60--200~s on a fourth-order mesh to maintain computational tractability over long simulation durations. Shorter-period structures (33--60~s) in RevealLO are inherited directly from the 1 h REVEAL baseline via mesh reprojection.
\end{enumerate}

%

\section{Conclusions}

In summary, the construction of RevealLO demonstrates that extending the inversion time window to 3.5 h is an effective, computationally tractable approach to overcoming the limitations of sparse receiver coverage in the southern hemisphere. By combining major-arc surface waves and multi-orbit body waves with careful artifact regularization and multi-scale mesh reprojection, RevealLO achieves a 15.6\% reduction of the L2 misfit on independent 3.5~hour validation data. RevealLO thus provides a more uniformly illuminated, data-consistent representation of the Earth's upper and mid-mantle, serving as an improved starting point for next-generation global tomographic models.

\section{Data and Code Availability}
\label{sec:data_availability}

RevealLO (provided in NetCDF format and as a Salvus-compatible HDF5 mesh) and LASIF 2.0 windowing configuration files are archived and openly available on Zenodo at \\
\href{https://doi.org/10.5281/zenodo.23646069}{https://doi.org/10.5281/zenodo.23646069}. The Salvus wave propagation software used for forward and adjoint simulations is available from \href{https://mondaic.com}{Mondaic Ltd.} (\url{https://mondaic.com}). All seismic waveform data used in this study were retrieved from public FDSN data centers and can be accessed via standard ObsPy clients.

\section*{Acknowledgements}

We acknowledge the Swiss National Supercomputing Centre (CSCS) for providing high-performance computing resources under grant allocation lp130, required to perform the global Salvus simulations. We thank the International Federation of Digital Seismograph Networks (FDSN) and its contributing network operators for maintaining and distributing the continuous broadband seismic waveform data used in this study.


\end{document}